\documentclass[showpacs,preprintnumbers,amsmath,amssymb]{revtex4}
\usepackage{graphicx}
\usepackage{dcolumn}
\usepackage{bm}
\begin{document}
\title{The He II Theory Preserving the Symmetry
of the Initial Hamiltonian of the System}
\author{I.M.Yurin}
 \email{yurinoffice@mail.ru}
\affiliation{%
I.M.Yurin, Fl.61, bld. 7, 22 Festivalnaya St, Moscow, 125581, Russian Federation}%

\date{\today}

\begin{abstract}
The article suggests a method for the construction of the number of
atoms preserving microscopic He II theory. The suggested theory can
provide the ground state wave function (WF) as an expansion in
series by small parameters. In addition, errors in the definition of
the WFs of the excited states turn out to be vanishingly small while
the system size increases. Predictions of the proposed theory and of
the Bogoliubov theory are identical if helium occupies a simply
connected volume. However, there are differences in the general
case. These differences are due to the fact that the operator of the
occupation number of the zero momentum state does not commute with
the creation and annihilation operators of phonons. This contradicts
the Bogoliubov assumption that the creation and annihilation
operators of atoms in the zero momentum state can be replaced by a
c-number. The latter, as is known, should commute with any operator.
This should lead to a difference in the results of calculation of
helium flow through the tunnel transition. Therefore, the suggested
theory is not equivalent to Bogoliubov theory.
\end{abstract}

\maketitle

\section{\label{sec:level1}Introduction}

In 1947 Bogoliubov suggested a microscopic description of He II
\cite{1} based on the Hamiltonian of two-particle interatomic
interaction:
\begin{equation}
\hat H = \sum\limits_{\bf q} {t_{\bf q} a_{\bf q}^ +  a_{\bf q} }  +
\sum\limits_{{\bf p},{\bf k},{\bf q}} {\frac{{V_{\bf q} }}{{2\Omega
}}a_{{\bf p} + {\bf q}}^ +  a_{{\bf k} - {\bf q}}^ +  a_{\bf k}
a_{\bf p} } ,
 \label{eq:thirteen}
\end{equation}
where $a_{\bf p}^ +$ and $a_{\bf p}$ are the creation and
annihilation operators for $He^4$ atoms with momentum ${\bf p}$, and
$\Omega$ is the system volume.

 The main idea of the calculation procedure suggested in this paper is the selection of
the two-particle interaction terms in the Hamiltonian  $\sim a_0^+
a_0$, $ a_0^+ a_0^+$, $a_0 a_0$ in order to consider the truncated
Hamiltonian
\begin{equation}
\hat H_{trunc}  = \sum\limits_{q > 0} {\hat h_\mathbf{q} } ,
\label{eq:one}
\end{equation}
while
\begin{equation}
\begin{array}{l}
 \hat h_{\bf q}  = t_{\bf q} a_{\bf q}^ +  a_{\bf q}  + \frac{{V_{\bf q} }}{{2\Omega }}a_{\bf q}^ +  a_{ - {\bf q}}^ +  a_0 a_0  + \frac{{V_{\bf q} }}{{2\Omega }}a_0^ +  a_0^ +  a_{\bf q} a_{ - {\bf q}}  + \frac{{V_{\bf q} }}{\Omega }a_0^ +  a_{\bf q}^ +  a_{\bf q} a_0 . \\
\label{eq:three_ref}
 \end{array}
\end{equation}
Hereinafter we will not consider the terms $\sim V_0$, because they
only add a constant to the energy of the system with a fixed number
of particles.

 Accounting for these terms only~(\ref{eq:three_ref}) in order to calculate the physical parameters
of the system is not in doubt, because the condensation of a
macroscopic number of atoms in a state with zero momentum results in
the fact that the transition matrix elements corresponding to the
selected terms are far superior in magnitude to the rest of the
matrix elements.

However, for the diagonalization procedure Bogoliubov replaced
operators $a_0^ +$ and $a_0$ with so-called c-numbers. As a result,
the truncated Hamiltonian, in contrast to the initial one, no longer
keeps the number of particles in the system, which means the breach
of the U(1) symmetry of the system. This state of affairs will
always be a concern in the scientific community \cite{2}, because it
is well known that the c-number hypothesis is still unproven
\cite{11}.

Previously, attempts were made to construct a microscopic He II
theory without breaking the U(1) symmetry \cite{9,3}, such as the
theory presented in this work. They all used the creation and
annihilation operators of phonons, which can be represented as
follows (cf. ~(\ref{eq:TwentyThree})):
\begin{equation}
\begin{gathered}
  \varphi _\mathbf{q}^ +   = u_\mathbf{q} a_\mathbf{q}^ +  a_0 \gamma _\mathbf{q}  + v_\mathbf{q} \gamma _\mathbf{q} a_0^ +  a_{ - \mathbf{q}} , \hfill \\
  \varphi _\mathbf{q}  = u_\mathbf{q} \gamma _\mathbf{q} a_0^ +  a_\mathbf{q}  + v_\mathbf{q} a_{ - \mathbf{q}}^ +  a_0 \gamma _\mathbf{q} , \hfill \\
\end{gathered}
\label{eq:DefPred}
\end{equation}
where
\begin{equation}
\begin{gathered}
  u_\mathbf{q}  = \frac{{\omega _\mathbf{q}  + t_\mathbf{q} }}
{{2\sqrt {t_\mathbf{q} \omega _\mathbf{q} } }}, \hfill \\
  v_\mathbf{q}  = \frac{{\omega _\mathbf{q}  - t_\mathbf{q} }}
{{2\sqrt {t_\mathbf{q} \omega _\mathbf{q} } }} \hfill \\
\end{gathered}
\label{eq:eleven_ref}
\end{equation}
with
\begin{equation}
\omega _{\bf q}  = \sqrt {t_{\bf q} \left( {t_{\bf q}  + 2V_{\bf q}
N/\Omega } \right)}.
\end{equation}
Parameters $u_{\bf q}$ and $v_\mathbf{q}$ coincide with parameters
of $u - v$ Bogoliubov's transformation.

 In paper \cite{3} $\gamma _{\bf q}$ were numerical values, but in paper \cite{9}
$\gamma _{\bf q}$ were an operator:
\begin{equation}
\gamma _{\bf q}  = \frac{1} {{\sqrt {\hat N} }},
\end{equation}
where $\hat N$ is the operator of the total number of atoms in the
system.

Both definitions  led to distortion of bosonic
operators~(\ref{eq:DefPred}), which resulted in a lack of precision
in calculating the WFs of the ground and excited states of
Hamiltonian $\hat H_{trunc}$. As a consequence, the authors \cite{9}
had to introduce small parameters irrelevant to the problem
resulting in further restrictions to the form of the WF of the
system ground state:
\begin{equation}
N - \left\langle {\hat n_0 } \right\rangle _0  \ll N,
\label{eq:condCastin}
\end{equation}
where $N$ is the number of atoms in the system, $\hat n_{\bf p}  =
a_{\bf p}^ +  a_{\bf p}$
 and $\left\langle {...} \right\rangle _0$
denotes averaging over the vacuum state of the phonon system $\Phi
_0$, which satisfies conditions
\begin{equation}
\varphi _{\bf q} \Phi _0  = 0.
\label{eq:condvacuum}
\end{equation}
The other author \cite{3} tried to avoid these limitations by
specifying numerically the WFs of the ground and excited states of
the system with the use of an approximation close to $\Phi
_0^{\left( 1 \right)}$ (see ~(\ref{eq:phi_0_1})) for the ground
state. This attempt was not successful, as expected. In particular,
the author obtained a gap in the phonon spectrum, which is contrary
to the Goldstone theorem \cite{4}.

This article suggests a new definition of the creation and
annihilation operators of phonons ~(\ref{eq:TwentyThree}), which
allows, on one hand, presenting the WF of the ground state of the
system in the form of expansion in series by small parameters and,
on the other hand, avoiding the loss of accuracy when constructing
the wave WFs of multiphonon states due to distortion of bosonic
operators.

\section{\label{sec:level1}Vacuum State of the Phonon System}
Let us consider the sequence of states $\Phi _0^{\left( n \right)}$
composed by the following rules:
\begin{equation}
 \begin{array}{l}
  \Phi _0^{\left( { - 1} \right)}  = 0, \\
  \Phi _0^{\left( 0 \right)}  = \frac{{\left[ {a_0^ +  } \right]^N }}{{\sqrt {N!} }}\left| 0 \right\rangle , \\
  \end{array}
 \end{equation}

\begin{equation}
\Phi _0^{\left( n \right)}  = \Phi _0^{\left( {n - 1} \right)}  +
\frac{1} {2}\sum\limits_{q_n  > 0} {\hat G\left( {\mathbf{q}_n }
\right)} \left( {\Phi _0^{\left( {n - 1} \right)}  - \Phi _0^{\left(
{n - 2} \right)} } \right),
\label{eq:phi_0_1}
\end{equation}
where
\begin{equation}
\hat G\left( {\bf q} \right) = \sum\limits_{2m \le \hat n_0 }
{\left( { - g\left( {\bf q} \right)} \right)^m \left[
{\frac{1}{{\sqrt {\hat n_{\bf q} } }}a_{\bf q}^ +  \frac{1}{{\sqrt
{\hat n_{ - {\bf q}} } }}a_{ - {\bf q}}^ +  } \right]^m \left[ {a_0
} \right]^{2m} \sqrt {\frac{{\left( {\hat n_0  - 2m}
\right)!}}{{\hat n_0 !}}} } ,
\end{equation}
\begin{equation}
 g\left( {\bf q} \right) = v_{\bf q} /u_{\bf q}  = {{\omega _{\bf q}
- t_{\bf q} } \over {\omega _{\bf q}  + t_{\bf q} }},
 \end{equation}
while $ {\bf q}_{i + 1}  \ne  \pm {\bf q}_i$, if $i > 0$, $n
\leqslant N/2$. It is easy to show that WFs $\Phi _0^{\left( n
\right)}$ are approximations to the WF of the phonon vacuum the more
accurate the larger the number $n$ is. For this purpose let us first
define creation and annihilation operators of phonons:
\begin{equation}
\begin{gathered}
  \varphi _\mathbf{q}^ +   = u_\mathbf{q} a_\mathbf{q}^ +  a_0 \frac{1}
{{\sqrt {\hat n_0 } }} + \frac{{v_\mathbf{q} }}
{{\sqrt {\hat n_0 } }}a_0^ +  a_{ - \mathbf{q}} , \hfill \\
  \varphi _\mathbf{q}  = \frac{{u_\mathbf{q} }}
{{\sqrt {\hat n_0 } }}a_0^ +  a_\mathbf{q}  + v_\mathbf{q} a_{ -
\mathbf{q}}^ +  a_0 \frac{1}
{{\sqrt {\hat n_0 } }}. \hfill \\
\end{gathered}
\label{eq:TwentyThree}
\end{equation}
Now it is possible to notice that adding terms $\sim \hat G\left(
{{\bf q}_n } \right)$ into ~(\ref{eq:phi_0_1}) enables compensating
WFs $\varphi _{\bf q} \left( {\Phi _0^{\left( {n - 1} \right)}  -
\Phi _0^{\left( {n - 2} \right)} } \right)$ that results from the
calculation of expression $\varphi _{\bf q} \Phi _0^{\left( n
\right)}$. Accordingly, $\varphi _{\bf q} \Phi _0^{\left( n
\right)}$ turns out to be zero with a precision of $ \sim
\prod\limits_{i = 1}^n {g\left( {{\bf q}_i } \right)}$ for $n > 0$.

The structure of WFs $\Phi _0^{\left( n \right)}$ makes it possible
to calculate the system parameters as a power series by small
parameters $g\left( {\bf q} \right)$. However, as will be shown
below, it is possible to avoid the calculation of complicated
expressions associated with the factorization into small parameters
$g\left( {\bf q} \right)$ by transition to operators $\varphi _{\bf
q}^ +$ and $\varphi _{\bf q}$ in the system description. These
operators can be now definitely interpreted as creation and
annihilation operators of phonons due to their commutators'
properties.

As a matter of fact, the commutation relations $\left[ {\varphi
_{\bf k}^ + ,\varphi _{\bf q}^ +  } \right]$ and $\left[ {\varphi
_{\bf k} ,\varphi _{\bf q}^ + } \right]$ are bosonic. This statement
should be understood in the sense that the action of the commutation
relations on the superposition of states, each of which contains at
least one atom with zero momentum gives zero for $\left[ {\varphi
_{\bf k}^ + ,\varphi _{\bf q}^ + } \right]$ and does not differ from
the action of the corresponding Kronecker delta for $\left[ {\varphi
_{\bf k} ,\varphi _{\bf q}^ + } \right]$ (cf. Appendix A).

On the other hand, it is easy to show that the fraction of states
not containing atoms with zero momentum for the state with
$N_\varphi$ phonons is $\sim \prod\limits_{2i \le N - N_\varphi  }
{g\left( {{\bf q}_i } \right)}$. It means that for large systems
with $N_\varphi  :N - N_\varphi \gg 1$ the commutation relations
under consideration become bosonic with any desired accuracy  on the
order of small parameters $g\left( {\bf q} \right)$: it is enough to
choose a sufficiently large $N -N_\varphi$. This allows determining
the orthonormal bases of states with any desired accuracy on the
order of small parameters $g\left( {\bf q} \right)$, if the number
of phonons is not too large.

In the future we shall use symbol $\Phi _0$ to designate the vacuum
state of the phonon system. For the construction of the
above-mentioned bases of states it is essential that
relationships~(\ref{eq:condvacuum}) be fulfilled with any desired
accuracy. On the other hand, for $\Phi _0^{\left( n \right)}$ these
relationships, as noted above, are true with a precision of $ \sim
\prod\limits_{i = 1}^{n} {g\left( {{\bf q}_i }\right)}$. Therefore,
relationships~(\ref{eq:condvacuum}) can be fulfilled with any degree
of accuracy: it is enough to choose a sufficiently large $N$.

Thus, the transition to the new creation and annihilation operators
of phonons~(\ref{eq:TwentyThree}) is equivalent to the calculation
of the WFs of the system with any desired accuracy on the order of
small parameters $g\left( {\bf q} \right)$, if the number of phonons
in the system $N_\varphi$ satisfies relationship $N - N_\varphi \gg
1$. The possibility of satisfying this relationship in the framework
of the developed theory without any loss of accuracy of the
calculated values is very important, specifically, for the
description of its thermodynamic properties.

\section{\label{sec:level1}Diagonalization of the Full Hamiltonian of the System}

In systems with very large expectation values of ${a_0^ +  a_0 }$ initial operators
$a_{\bf q}^+$ and $a_{\bf q}$ can be expressed in terms of operators $\varphi _{\bf q}^ +$ and $\varphi _{\bf q}$
as follows:
\begin{equation}
\begin{gathered}
a_{\bf q}^ +   = \frac{1}{{\sqrt {\hat n_0 } }}a_0^ +  \left(
{u_{\bf q} \varphi _{\bf q}^ +   - v_{\bf q} \varphi _{ - {\bf q}} }
\right), \hfill \\
a_{\bf q}  = \left( {u_{\bf q} \varphi _{\bf q}  - v_{\bf q} \varphi
_{ - {\bf q}}^ +  } \right)a_0 \frac{1}{{\sqrt {\hat n_0 } }}. \hfill \\
\end{gathered}
\label{eq:TwentySix}
\end{equation}

This definition should be understood in the sense that the action of
the parent operators $a_\mathbf{q}^ +$ and $a_\mathbf{q}$ on the
state containing at least one atom with zero momentum does not
differ from the action of the operators taken from the rhs of
~(\ref{eq:TwentySix}). So, ~(\ref{eq:TwentySix}) can be used to
replace the parent operators with the phonon ones for large systems in case when it
comes to states with a not very large number of phonons. The
reasoning here is a little different from the above reasoning on commutation
relations $\left[ {\varphi _{\bf k}^ +  ,\varphi _{\bf q}^ +  }
\right]$ and $\left[ {\varphi _{\bf k} ,\varphi _{\bf q}^ +  }
\right]$.

However, the operator notation of the Hamiltonian
~(\ref{eq:thirteen}) is different from the representation of the
Hamiltonian in the creation and annihilation operators of the
so-called bogolons, if the creation and annihilation operators of
the bogolons are replaced with operators $\varphi _{\bf q}^ +$ and
$\varphi _{\bf q}$, respectively, even if one takes into account
commutativity of operators $\varphi _{\bf q}^ +$ and $\varphi _{\bf
q}$ with operators $\frac{1}{{\sqrt {\hat n_0 } }}a_0^ +$ and $a_0
\frac{1}{{\sqrt {\hat n_0 } }}$ (Appendix B). This is due to the
absence of a clear procedure for calculating the c-number in the
Bogoliubov theory.

In this connection let us dwell upon the calculation of the phonon
spectrum restricting ourselves to a simplest basis of four states:
$\Phi _0$, $\varphi _{\bf q}^ + \Phi _0$, $\varphi _{ - {\bf q}}^ +
\Phi _0$ and $\varphi _{\bf q}^ + \varphi _{ - {\bf q}}^ +  \Phi
_0$. The matrix elements of the Hamiltonian for the transitions
connecting these states determine the effective Hamiltonian bilinear
by the creation and annihilation operators of phonons.
Diagonalization of this Hamiltonian leads to the following
expression for the spectrum dispersion:
\begin{equation}
\tilde \omega _q  = \sqrt {t_q \left( {t_q  + 2V_q \left\langle
{\hat n_0 } \right\rangle _0 /\Omega } \right)} \text{.}
\label{eq:TwentySeven_ref}
\end{equation}
Calculating the above matrix elements became possible, because
operator $\hat n_0$ behaves as a number upon affecting $\Phi _0$, if
$N \gg 1$. Indeed, taking into account that
\begin{equation}
\hat n_0 \Phi _0  = \left( {N - \sum\limits_{{\bf q} \ne 0} {\hat
n_{\bf q} } } \right)\Phi _0 ,
\end{equation}
\begin{equation}
\hat n_0 ^2 \Phi _0  = \left( {N - \sum\limits_{{\bf q'} \ne 0}
{n_{{\bf q'}} } } \right)\left( {N - \sum\limits_{{\bf q} \ne 0}
{n_{\bf q} } } \right)\Phi _0
\end{equation}
it is easy to evaluate the operator $\hat n_0$ mean root square
deviation from its average value
\begin{equation}
\delta n_0  = \sqrt {\left\langle {\hat n_0 } \right\rangle _0^2  -
\left\langle {\hat n_0^2 } \right\rangle _0 }  \sim \sqrt
{\left\langle {\hat n_0 } \right\rangle _0 } .
\end{equation}
So, $\delta n_0 /\left\langle {\hat n_0 } \right\rangle _0  \sim
\left\langle {\hat n_0 } \right\rangle _0^{ - 1/2}$. This means
that, when $\left\langle {\hat n_0 } \right\rangle _0  \gg 1$ ,
operator $\hat n_0$ behaves as number $\left\langle {\hat n_0 }
\right\rangle _0$ upon affecting vacuum state $\Phi _0$, which is in
agreement with Bogoliubov c-number hypothesis.

 At the same time it is important to note that for the realistic
interatomic potential $V_q$ we have $\left\langle {\hat n_0 }
\right\rangle _0  \ll N$, which leads to a considerable difference
in the dispersion estimates in the suggested theory and the
Bogoliubov approach. At this point it is important to note that in
our reasoning we did not have to impose restrictions of the
type~(\ref{eq:condCastin}) on the WFs (as a rule, these restrictions
are used by the authors when considering the He II system
\cite{2,8,9}). This, apparently, makes possible the in fact
first-principle calculation of the spectrum with a realistic
interatomic potential.

Note now that, when we searched the vacuum state of the phonon
system, parameters $u_{\bf q}$ and $v_{\bf q}$ in section 2
essentially turned out not to be related to definition
~(\ref{eq:eleven_ref}). For these parameters there is only the
requirement related to the need to fulfill the boson commutation
relations:
\begin{equation}
u_{\bf q}^2  - v_{\bf q}^2  = 1.
\end{equation}
In such a situation it is advisable to redetermine them:
\begin{equation}
\begin{array}{l}
 u_{\bf q}  = \frac{{\tilde \omega _{\bf q}  + t_{\bf q} }}{{2\sqrt {t_{\bf q} \tilde \omega _{\bf q} } }}, \\
 v_{\bf q}  = \frac{{\tilde \omega _{\bf q}  - t_{\bf q} }}{{2\sqrt {t_{\bf q} \tilde \omega _{\bf q} } }}. \\
 \end{array}
 \label{eq:TwentyNine_ref}
\end{equation}
It remains only to determine the value of $\left\langle {\hat n_0 }
\right\rangle _0$. In order to do this, let us use the formula
relating to the conservation of the number of atoms in the system:
\begin{equation}
\left\langle {\hat n_0 } \right\rangle _0  + \sum\limits_{{\bf q}
\ne 0} {\left\langle {\hat n_{\bf q} } \right\rangle _0 }  = N.
\end{equation}
After this, using ~(\ref{eq:TwentySix}) we have:
\begin{equation}
\left\langle {\hat n_0 } \right\rangle _0  = N - \sum\limits_{{\bf
q} \ne 0} {v_{\bf q}^2 }.
\label{eq:ThirtyOne_ref}
\end{equation}
Equations ~(\ref{eq:TwentySeven_ref}, \ref{eq:TwentyNine_ref},
\ref{eq:ThirtyOne_ref}) form a self-consistent system for
determining the phonon spectrum. The next important step in the
calculation of the phonon spectrum of the system is the
consideration of the trilinear terms of the interaction Hamiltonian
by operators $\varphi _{\bf q}^ +$ and $\varphi _{\bf q}$. This
should be the theme of a separate article.

\section{\label{sec:level1}Differences from the Bogoliubov Theory}
The above approach to the calculation of the phonon spectrum
indicated the proximity of the presented theory to the Bogoliubov
theory. However, as will be shown below, there are differences. This
can easily be seen by considering, for example, commutator
\begin{equation}
\left[ {a_0^ +  a_0 ,\varphi _{\bf q}^ +  } \right] =  - u_{\bf q}
a_{\bf q}^ +  a_0 \frac{1} {{\sqrt {\hat n_0 } }} + \frac{{v_{\bf q}
}} {{\sqrt {\hat n_0 } }}a_0^ +  a_{ - {\bf q}}  \ne 0.
\end{equation}
Obviously, in this case $a_0^ +  a_0$ does not behave as a c-number
squared, which contradicts the Bogoliubov theory.

This fact, in particular, results in a difference in the motion
equations for the creation and annihilation operators of phonons in
the physically important case when the system interacts with
potential $\mu \left( t \right)$ constant in the space. In this case
the Hamiltonian includes an additional term
\begin{equation}
\hat U = \mu \left( t \right)\hat N.
\end{equation}
Term $\hat U$ commutes with operator $\varphi _{\bf q}$, in contrast
to the annihilation operator of the bogolon
\begin{equation}
b_{\bf q}  = u_{\bf q} a_{\bf q}  + v_{\bf q} a_{ - {\bf q}}^ +  ,
\end{equation}
for which we have
\begin{equation}
\left[ {\hat U,b_{\bf q} } \right] = \mu \left( t \right)\left( { -
u_{\bf q} a_{\bf q}  + v_{\bf q} a_{ - {\bf q}}^ +  } \right) \ne 0.
\end{equation}
This difference leads to a difference in the motion equations for
the annihilation operators of bogolons and operators $\varphi _{\bf
q}$. Thus, parameter $u_{\bf q}$ in the Bogoliubov theory acquires
phase factor $\exp \left( {i\theta } \right)$, and $ v_{\bf q}$,
accordingly, acquires factor $\exp \left( { - i\theta } \right)$.
Besides,
\begin{equation}
\dot \theta  = \mu \left( t \right).
\end{equation}
In the presented theory, however, parameters $u_{\bf q}$ and $v_{\bf
q}$ do not acquire phase factors, which leads to a difference in the
results of the consideration of problems related to the Josephson
effect.

In this respect let us now consider the following situation. Let us
bring into contact two systems of helium with the use of Hamiltonian
\begin{equation}
\hat H_T  = \sum\limits_{{\bf k},{\bf q}} {T_{{\bf k},{\bf q}}
a_{\bf k}^ +  b_{\bf q} }  + H.c.
\label{eq:tunnelHam}
\end{equation}
We assume that the operators corresponding to the momentum ${\bf k}$
are related to the atoms of the first system, and that those
corresponding to the momentum ${\bf q}$  are related to the atoms of
the second system. For simplicity, we assume
\begin{equation}
T_{0,{\bf q}}  = T_{{\bf k},0}  = T_{0,0}  = 0
\end{equation}
in order to avoid description of transitions with the participation
of atoms with zero momentum in two systems. Let us note at this
point that the Hamiltonian of the type~(\ref{eq:tunnelHam}) is
associated with helium atoms tunneling through solid membranes. At
the same time, the penetration in this case is so small that its
experimental study is not possible at the present time. Researchers
use perforated membranes, and it is assumed that in this case the
exchange of atoms between the 2 systems is not described by
Hamiltonian~(\ref{eq:tunnelHam}). From this point of view the
consideration presented below is of purely methodological value in
order to compare the results with analogous results in the BCS
theory.

Let us apply transformation $\exp \left( {i\theta \hat N} \right)$
in the space of basis states of one of the systems, that is, let us
replace parameters $\left( {u_{\bf q} ,v_{\bf q} } \right)$ with
$\left( {u_{\bf q} \exp \left( {i\theta } \right),v_{\bf q} \exp
\left( { - i\theta } \right)} \right)$. It would seem that the
observed values of the system should not be changed according to the
principles of quantum physics. However, an elementary calculation
shows the dependence on $\theta$, for example, of the second order
energy perturbation
\begin{equation}
E^{\left( 2 \right)}  =  - \sum\limits_{{\bf k},{\bf q}} {\left|
{T_{{\bf k},{\bf q}} } \right|^2 \frac{{\left| {u_{\bf k} v_{\bf q}
+ u_{\bf q} v_{\bf k} } \right|^2 }} {{\omega _{\bf k}  + \omega
_{\bf q} }}}
\end{equation}
caused by the transition Hamiltonian~(\ref{eq:tunnelHam}).

In contrast to independent researchers \cite{6}, BCS theory
supporters do not see any paradox in the dependence of $E^{\left( 2
\right)}$ on $\theta$ (cf. Eq.(40) and Eq.(3) in \cite{10}). They
even use it to calculate the Josephson current (cf. Eq. (10) in
\cite{10}). Using a system of reasoning conventional in the BCS
theory, an expression for Josephson flux $F$ can be obtained:
\begin{equation}
F = \frac{{dE^{\left( 2 \right)} }} {{d\theta }},
\label{eq:Flow}
\end{equation}
since in the absence of dissipation flow $F$ is correlated to the
system energy $E$ by the expression
\begin{equation}
F\mu \left( t \right)dt = Fd\theta  = dE.
\end{equation}
Of course, there is no term ~(\ref{eq:Flow}) in the expression for
flow in the theory that is being developed, because there are no
corresponding phase factors in parameters $u_{\bf q}$ and $v_{\bf
q}$. This is just what proves the non-equivalence of this theory to
the Bogoliubov theory. Moreover, the absence of such terms in the
presented strict He II theory casts doubt both on the physical
status of expressions for Josephson current in the BCS theory
resulting, as well as expression~(\ref{eq:Flow}), exclusively due to
the breach of the U(1) symmetry, and on the BCS theory itself. At
this point it is appropriate to note that, apparently, the only
superconductivity theory which currently explains consistently the
Josephson effect \cite{6} has essentially nothing to do with the
basics of the BCS theory. Presumably, the results of the experiment
suggested in paper \cite{12} will help in finding the correct theory
of superconductivity.

 \section{\label{sec:level1}Conclusion}
 In this paper a microscopic foundation of the He II theory preserving the U(1) symmetry of the initial
 Hamiltonian was created. If we consider the case of a simply-connected system
 of helium, the developed theory, apparently, is equivalent to the Bogoliubov theory,
 if the latter includes the procedure for calculating the c-number based on the results of this work.
 However, in the general case, the suggested theory is not equivalent to the
 Bogoliubov theory, because operator $a_0^ +  a_0$  does not commute with the
 creation and annihilation operators of phonons.

Despite the equivalence of the approaches in the case of
simply-connected systems, the obtained results indicate the need for
a major revision of the He II theory as a whole. In fact, the
central assumption of this theory is the so-called two-fluid model,
while the density of the normal component $\rho _n$ and that of the
superfluid component $\rho _s$ being related to the system density
$\rho$ by a naive balance condition $\rho = \rho _n  + \rho _s $
\cite{13}. Besides, the normal component is associated with the
phonon system. So, the fact that the operator of the total number of
atoms commutes with operators $\varphi _{\bf q}^ +$ and $\varphi
_{\bf q}$ directly contradicts this assumption, because mass
operator $\hat M \sim \hat N$.

In this respect the author makes the assumption that the
disappearance of superfluidity at the transition point is due to
zeroing of the phonon velocity in the long wave limit rather than to
$\rho _n$ zeroing, as is thought in the presently recognized He II
theory. Obviously, according to the suggested conception the
velocity of phonons as quantum objects is not necessarily associated
to the compressibility of helium, at least at temperatures close to
the transition temperature. Calculating the influence of thermally
excited phonons on their spectrum considering the phonon-phonon
interaction can ascertain the truth in this matter. For obtaining
these fundamental results it is important to avoid relations of the
type~(\ref{eq:condCastin}), so that their reliability would not be
in doubt.

The suggested theory makes possible such a calculation. However, in
the Bogoliubov theory, even if the calculation were carried out, its
reliability would be in doubt due to lack of proof for the c-numbers
hypothesis. In this regard it may be noted that the conclusion about
the necessity of revising the He II theory could probably have been
made on the basis of the Bogoliubov theory. However, the lack of
proof for the c-numbers hypothesis made it impossible to indicate
reliably the mismatch of the microscopic and phenomenological parts
of the He II theory.

On the other hand, the mass operator in the Bogolyubov theory does
not commute with the creation operators of bogolons. So, the
conclusion about the unsatisfactory state of the theory of He II as
a whole could be made only on the basis of a rigorous and objective
analysis, which, unfortunately, has not been performed. Apparently,
this was also due to the statement by Bogolyubov himself \cite{1}
that his theory is a very crude method of the system examination.
Accordingly, the discrepancies in comparison with the
phenomenological theory in such a situation seemed to be quite
acceptable.

\begin{acknowledgments}
The author is grateful to Professor A.A.  Rukhadze for the moral
support of author's works on theories of superphenomena. In
addition, the author thanks D.Sc.  S.A. Trigger for criticism of
author's first attempts to solve the problem, which were
subsequently refused.
\end{acknowledgments}

\appendix
\section{Calculation of commutators $\left[ {\varphi _{\bf k}^ +  ,\varphi _{\bf q}^ +  } \right]$ and $\left[ {\varphi _{\bf k} ,\varphi _{\bf q}^ +  } \right]$}
Direct substitution of ~(\ref{eq:TwentyThree}) into $\left[ {\varphi
_{\bf k}^ + ,\varphi _{\bf q}^ +  } \right]$
  gives
\begin{equation}
\begin{array}{l}
 \left[ {\varphi _{\bf k}^ +  ,\varphi _{\bf q}^ +  } \right] = u_{\bf k} u_{\bf q} \left[ {a_{\bf k}^ +  ,a_{\bf q}^ +  } \right]a_0 \frac{1}{{\sqrt {\hat n_0 } }}a_0 \frac{1}{{\sqrt {\hat n_0 } }} + v_{\bf k} v_{\bf q} \left[ {a_{ - {\bf k}} ,a_{ - {\bf q}} } \right]\frac{1}{{\sqrt {\hat n_0 } }}a_0^ +  \frac{1}{{\sqrt {\hat n_0 } }}a_0^ +   \\
  + u_{\bf k} v_{\bf q} \left( {a_{\bf k}^ +  a_{ - {\bf q}} \frac{1}{{\hat n_0  + 1}}a_0 a_0^ +   - a_{ - {\bf q}} a_{\bf k}^ +  \frac{1}{{\hat n_0 }}a_0^ +  a_0 } \right) + u_{\bf q} v_{\bf k} \left( {a_{ - {\bf k}} a_{\bf q}^ +  \frac{1}{{\hat n_0 }}a_0^ +  a_0  - a_{\bf q}^ +  a_{ - {\bf k}} \frac{1}{{\hat n_0  + 1}}a_0 a_0^ +  } \right). \\
 \end{array}
\end{equation}
Considering the action of operator $\left[ {\varphi _{\bf k}^ +
,\varphi _{\bf q}^ +  } \right]$ on the superposition of states
$\left| S \right\rangle$, each of which contains at least one atom
with zero momentum, we have
\begin{equation}
\left[ {\varphi _{\bf k}^ +  ,\varphi _{\bf q}^ +  } \right]\left| S
\right\rangle  = u_{\bf k} v_{\bf q} \left[ {a_{\bf k}^ +  ,a_{ -
{\bf q}} } \right]\left| S \right\rangle  + u_{\bf q} v_{\bf k}
\left[ {a_{ - {\bf k}} ,a_{\bf q}^ +  } \right]\left| S
\right\rangle  = 0.
\end{equation}
Analogously, for $\left[ {\varphi _{\bf k} ,\varphi _{\bf q}^ +  }
\right]$ we have
\begin{equation}
\begin{array}{l}
 \left[ {\varphi _{\bf k} ,\varphi _{\bf q}^ +  } \right] = u_{\bf k} v_{\bf q} \left[ {a_{\bf k} ,a_{ - {\bf q}} } \right]\frac{1}{{\sqrt {\hat n_0 } }}a_0^ +  \frac{1}{{\sqrt {\hat n_0 } }}a_0^ +   + u_{\bf q} v_{\bf k} \left[ {a_{ - {\bf k}}^ +  ,a_{\bf q}^ +  } \right]a_0 \frac{1}{{\sqrt {\hat n_0 } }}a_0 \frac{1}{{\sqrt {\hat n_0 } }} \\
  + u_{\bf k} u_{\bf q} \left( {a_{\bf k} a_{\bf q}^ +  \frac{1}{{\hat n_0 }}a_0^ +  a_0  - a_{\bf q}^ +  a_{\bf k} \frac{1}{{\hat n_0  + 1}}a_0 a_0^ +  } \right) +  \\
  + v_{\bf k} v_{\bf q} \left( {a_{ - {\bf k}}^ +  a_{ - {\bf q}} \frac{1}{{\hat n_0  + 1}}a_0 a_0^ +   - a_{ - {\bf q}} a_{ - {\bf k}}^ +  \frac{1}{{\hat n_0 }}a_0^ +  a_0 } \right). \\
 \end{array}
\end{equation}
Considering the action of operator $\left[ {\varphi _{\bf k}
,\varphi _{\bf q}^ +  } \right]$ on the superposition of states
$\left| S \right\rangle$, each of which contains at least one atom
with zero momentum, we have
\begin{equation}
\left[ {\varphi _{\bf k} ,\varphi _{\bf q}^ +  } \right]\left| S
\right\rangle  = u_{\bf k} u_{\bf q} \left[ {a_{\bf k} ,a_{\bf q}^ +
} \right]\left| S \right\rangle  + v_{\bf k} v_{\bf q} \left[ {a_{ -
{\bf k}}^ +  ,a_{ - {\bf q}} } \right]\left| S \right\rangle  =
\delta _{\bf q}^{\bf k} \left| S \right\rangle ,
\end{equation}
where $\delta _{\bf q}^{\bf k}$ is the 3D Kronecker delta.

Thus, commutators $\left[ {\varphi _{\bf k}^ +  ,\varphi _{\bf q}^ +
} \right]$ and $\left[ {\varphi _{\bf k} ,\varphi _{\bf q}^ +  }
\right]$ are not different from ordinary bosonic commutators in
their action on the superposition of states, each of which contains
at least one atom with zero momentum.

\section{Commutativity of operators $\varphi _{\bf q}^ +$
  and $\varphi _{\bf q}$ with operators
$\frac{1}{{\sqrt {\hat n_0 } }}a_0^ + $ and $ a_0 \frac{1}{{\sqrt
{\hat n_0 } }}$}

For example, let us calculate commutator $\left[ {\varphi _{\bf q}^
+  ,\frac{1}{{\sqrt {\hat n_0 } }}a_0^ +  } \right]$:

\begin{equation}
\left[ {\varphi _{\bf q}^ +  ,\frac{1}{{\sqrt {\hat n_0 } }}a_0^ + }
\right] = u_{\bf q} a_{\bf q}^ +  \left[ {a_0 \frac{1}{{\sqrt {\hat
n_0 } }},\frac{1}{{\sqrt {\hat n_0 } }}a_0^ +  } \right] = u_{\bf q}
a_{\bf q}^ +  \left( {\frac{1}{{\hat n_0  + 1}}a_0 a_0^ + - 1}
\right) = 0.
\end{equation}
The commutators $\left[ {\varphi _{\bf q}^ +  ,a_0 \frac{1}{{\sqrt
{\hat n_0 } }}} \right]$,
 $\left[ {\varphi _{\bf q} ,\frac{1}{{\sqrt {\hat n_0 } }}a_0^ +  }
\right]$ and $\left[ {\varphi _{\bf q} ,a_0 \frac{1}{{\sqrt {\hat
n_0 } }}} \right]$ can be considered analogously.

\bibliography{Comparison}

\end{document}